\documentclass{appolb}
\usepackage{epsfig}
\usepackage{amsmath}
\begin{document}
% \eqsec  % uncomment this line to get equations numbered by (sec.num)
\title{Automated resummation of jet observables in QCD%
\thanks{Talk presented at the XXXIII International Symposium
on
Multiparticle Dynamics,
September 5-11, 2003,
Krak\'ow, Poland.}%
% you can use '\\' to break lines
}
\author{Andrea Banfi
\address{NIKHEF Theory group\\
P.O. Box 41882 1009 DB Amsterdam, the Netherlands\\
E-mail: andrea.banfi@nikhef.nl}
}
\maketitle
\begin{abstract}
  We build a computer code that fully automates the resummation of
  jet-observable distributions at next-to-leading logarithmic
  accuracy. As an application we present results for a jet shape in
  hadronic dijet production.
\end{abstract}
  
\section{Introduction}
Understanding the physics of event shapes and jet rates (collectively
{\em jet observables}) requires a deep knowledge of QCD dynamics.
Their distributions (defined as the fraction of events $\Sigma(v)$
with the observable's value $V$ less than $v$) explore a wide range of
physical scales, from the more inclusive region $v={\cal O}(1)$, well
described by fixed order perturbative (PT) calculations, to the
extreme exclusive region $v \to 0$ where the language of quarks and
gluons is no longer applicable.
Describing the intermediate region
requires a resummation of logarithmic enhanced contributions that
appear at all orders in the PT expansion \cite{CTTW}. 
Resummation can then be seen as the link between PT and NP physics.
Given that interest, a number of event shapes have been introduced,
and for most of them there exist both next-to-leading order (NLO)
calculations as well as resummed predictions at next to-leading
logarithmic (NLL) accuracy. 

Unfortunately, while it is straightforward to compute a NLO
distribution by just interfacing a fixed order Monte Carlo program
with the observable definition in the form of a computer
routine, for what concerns resummation a separate analytic
calculation has to be performed for each observable.%~\cite{resum}.

Together with Gavin Salam and Giulia Zanderighi we started then a
project whose goal is the construction of a computer code that
takes as input from the user the
definition of an observable and returns its resummed distribution at NLL
accuracy.  The starting point of such a programme is the
classification of LL and NLL effects.

\section{Automated resummation}
\label{sec:auto}
Consider a Born system consisting of $n$ hard partons
(\emph{legs}) $\{p_i\}_{i=1,\ldots,n}$ and an observable $V$ that vanish in
the Born limit, i.e. $V(\{p\})=0$.

For most jet observables LL contributions can be addressed by
computing $\Sigma(v)$ when a single soft gluon collinear to the
hard legs is emitted and exponentiating the result. 

For any of these \emph{exponentiating} quantities, NL logarithms can
have different origins:
\begin{itemize}
\item emission of gluons soft at large angles or hard and
collinear. Such contributions 
can be resummed by exponentiating the single emission result for
$\Sigma(v)$;
\item {\it multiple emission}
effects, which originate when, given a set of emitted partons
$\{k_i\}$, all $V(k_i) < v$
but $V(\{k_i\}) > v$ (or viceversa), so that one has to take
into account how all emissions contribute to build up the observable's
value;
\item {\it non-global} logarithms~\cite{DS}, that arise whenever $V$ is
  sensitive to emissions only in a part of the phase space.
  Unfortunately such NLL contributions are known only in the large
  $N_c$ limit, thus reducing the accuracy of resummed predictions for
  non-global observables.
\end{itemize}

Given the above classification we restrict ourselves to exponentiating
global observables.  Under the additional hypothesis that, after a soft
emission collinear to leg $\ell$, the observable's dependence on the
emitted transverse momentum $k_t$, rapidity $\eta$ and azimuthal angle
$\phi$ (all with respect to $\ell$) obeys the following
parametrisation
\begin{equation}
  \label{eq:SC}
  V(k) = d_{\ell}\left(\frac{k_t}{Q}\right)^{a_\ell} 
  e^{-b_\ell\eta} g_\ell(\phi)\>, 
\end{equation}
at NLL accuracy $\Sigma(v)$ is given by the master formula~\cite{BSZ}
\begin{equation}
  \label{eq:master}
  \Sigma(v) = e^{-R(v)} {\cal F}(R'(v))\>,\qquad 
  R'(v) = -v\frac{dR(v)}{dv}\>.
\end{equation}
Here $R(v)$ embodies all LL contributions and the NLL contributions
that can be taken into account by exponentiating the single emission
result. This function, computed analytically, depends parametrically
on  $a_\ell$, $b_\ell$, $d_\ell$ as well as on the
azimuthal average $\langle \ln g_{\ell}(\phi)\rangle$.

The NLL function ${\cal F}(R')$ represents the multiple emission
correction factor. Its general expression is \cite{numsum}
\begin{equation}
  \label{eq:F-def}
  {\cal F}(R')=\left\langle 
  \exp\left\{-R'\ln\frac{V(k_1,\dots,k_n)}{\max\{V(k_1),\dots,V(k_n)\}}
  \right\}
\right\rangle\>,
\end{equation}
where the average is taken over all configuration of soft and
collinear gluons such that, for fixed $R'$, the probability density of
emissions along leg $\ell$ is constant (the actual value of the
constant depends on $R'$
and on the colour factors of the hard legs, see \cite{BSZ}).

Although the above procedure is rather complicated, 
we were able to embody it a computer code, CAESAR, that completely
automates the resummation of any exponentiating global
jet-observable~\cite{BSZ}.

\section{Event-shapes in hadronic dijet production}
\label{sec:hhevs}
Let us show how the program works in the specific case of an event shape in
hadronic dijet production. Given a unit vector $\vec n$ orthogonal to
the beam axis we define the global transverse thrust:
\begin{equation}
  \label{eq:thrust}
  \tau_{t,g} = 1-\max_{\vec n}\frac{\sum_i |\vec p_{ti} \cdot \vec n|}
  {\sum_i |\vec p_{ti}|}\>,
\end{equation}
where $\vec p_{ti}$ are the final state transverse momenta with respect to
the beam.

The program chooses a reference Born event with two incoming and two
outgoing hard partons and generates soft and collinear emissions. It
first checks that a resummation in the $(2+2)$-jet limit is feasible
and that the observable is global. It then verifies that the
parametrisation \eqref{eq:SC} holds and determines for each leg
$a_\ell, b_\ell, d_\ell$. It recognises also whether
$g_\ell(\phi)=|\sin\phi|^{c_\ell}$, with $c_\ell$ integer, otherwise
tabulates the azimuthal dependence.  The program then should determine
whether the observable exponentiates and compute the function ${\cal
  F}(R')$ via a Monte Carlo procedure \cite{numsum}. In this case it
recognises that $\tau_{t,g}$ is additive, that is
$V(k_1,\dots,k_n)=V(k_1)+\dots+V(k_n)$, which allows us to skip all
the subsequent steps, since for additive observables exponentiation
holds and ${\cal F}(R')=e^{-\gamma_E R'}/\Gamma(1+R')$ \cite{CTTW}.
  
\begin{figure}[htbp]
  \begin{minipage}[l]{.4\textwidth}
{\small
       \begin{tabular}{| l | c | c|}
      \hline
      Test  & result \\
      \hline
      check number of jets &    T  \\ 
      globalness  &  T  \\ 
      additivity   &  T  \\ 
      exponentiation             &  T  \\ 
      \hline
    \end{tabular}    

 \begin{tabular}{| c | c | c | c | c | c |}
 \hline
 $\ell$ & $a_{\ell}$ & $b_{\ell}$ & $g_{\ell}(\phi)$ & $d_{\ell}$  \\
 \hline
 \hline
1 & 1 & $ 0$ & tabulated &  1.02062    \\ 
 \hline                               
2 & 1 & $ 0$ & tabulated &  1.02062    \\ 
 \hline                               
3 & 1 & $ 1$ &$\sin^2\phi$ &  1.04167  \\ 
 \hline                               
4 & 1 & $ 1$ &$\sin^2\phi$ &  1.04167  \\ 
 \hline
 \end{tabular} 
}
  \end{minipage}
  \begin{minipage}[r]{.6\textwidth}
    \begin{flushright}
      \epsfig{file=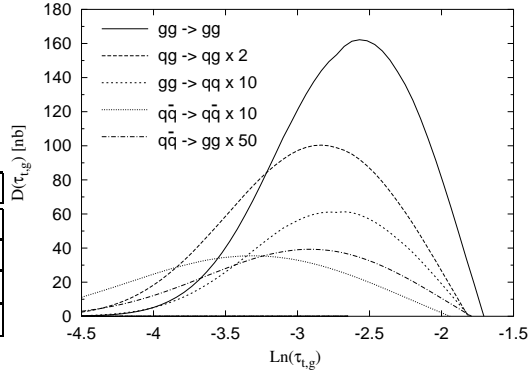, width=.95\textwidth}
    \end{flushright}
  \end{minipage}
  \caption{A sample output of CAESAR: the results of the observable analysis (left)
    and the resummed distribution at NLL accuracy (right) for the
    global transverse thrust.  The analysis has been performed with a
    reference Born configuration with the two outgoing partons
    back-to-back at an angle $\cos\theta=0.2$ with respect to the
    beam. The final result does not depend on the chosen
    configuration, as the program can check automatically.  }
  \label{fig:result}
\end{figure}

These results are automatically tabulated (see fig.~\ref{fig:result}),
and used as inputs for the master formula \eqref{eq:master} to
compute the NLL resummed $\tau_{t,g}$ distribution. In
fig.~\ref{fig:result} we see the differential distribution
$D(\tau_{t,g})=d\Sigma(\tau_{t,g})/d\ln\tau_{t,g}$ at the Tevatron run
II c.o.m. energy $\sqrt s = 1.96\mbox{TeV}$ corresponding to selected
dijets with $E_t>50 \mbox{GeV}$ and $|\eta|<1$. We use the
CTEQ6M pdf set~\cite{CTEQ} and set factorisation and renormalisation
scales at the partonic c.o.m.  energy. The plot shows a clean
separation among the various partonic channels, information that can
be exploited for fits of the parton densities.

\section{Conclusions and outlook}
\label{sec:fine}
The main feature of our approach is that the output curves are pure
NLL functions, without contamination with spurious subleading
contributions, so that they can
be straightforwardly matched with fixed order results, and any
hadronisation model can be applied.
At present our effort is concentrated in releasing the first
version of the program. 
The next step will be including in the code an
automated matching with NLO calculations \cite{NLO}, 
which will open the way to a vast amount of phenomenological analyses.

{\bf Acknowledgments.}
My wife and I are really grateful to the organisers for the warm
hospitality and the pleasant atmosphere during the whole conference.

\end{document}